\documentclass[preprint,preprintnumbers,nofootinbib,aps]{revtex4-2}
\usepackage[english]{babel}
\usepackage{color}
\usepackage{amsmath}
\usepackage{physics}
\usepackage{soul}
\usepackage{xr-hyper}
\usepackage[hidelinks]{hyperref}
\usepackage{diagbox}
\usepackage{multirow}
\usepackage{amsfonts}
\usepackage{amssymb}
\usepackage{latexsym}
\usepackage{graphicx}
\usepackage{adjustbox}
\usepackage{xr}
\usepackage{caption}
\usepackage{enumitem}
\usepackage{subfigure}
\usepackage{seqsplit}
\usepackage{ytableau}
\usepackage{booktabs}
\usepackage{mathtools}
\usepackage{cancel}

\newcommand{\pa}{\partial}

\newcommand{\rar}{\rightarrow}

\newcommand{\ii}{\mathrm{i}}
\newcommand{\e}{\mathrm{e}}
\numberwithin{equation}{section}

\begin{document}
\title{Time-Dependent Logarithmic Perturbation Theory\\
	for Quantum Dynamics: Formulation and Applications}
	
\author{J.C.~del~Valle$^{1}$}
\thanks{Corresponding author}
\email{delvalle@fisica.unam.mx}
\affiliation{$^1$Instituto de F\'isica, Universidad~Nacional~Aut\'onoma~de~M\'exico, Ciudad de M\'exico 04510, Mexico}

\author{P.~Bergold$^{2}$}
\affiliation{$^2$Instytut Matematyki Stosowanej, Politechnika Gda\'nska, 80--233~Gda\'nsk, Poland}

\author{K.~Kropielnicka$^{3}$}
\affiliation{$^3$Instytut Matematyki, Uniwersytet Gda\'nski, 80--308 Gda\'nsk, Poland}

\begin{abstract}
	We present the time-dependent extension of logarithmic perturbation theory (LPT) for non-relativistic quantum dynamics governed by the Schr\"odinger equation.
	Assuming that the system is initially in a nondegenerate ground state, the logarithm of the wave function is expanded in powers of a coupling constant.
	The resulting hierarchical equations for the perturbative corrections are defined via a gauge-rotated Hamiltonian of the unperturbed system and lead to closed-integral expressions for the time-dependent corrections.
	This integral structure, the hallmark of time-independent LPT, is preserved in the present extension. 
	Furthermore, dynamic energy shifts arise naturally within this framework in the form of time-averaged expectation values of pseudopotentials and are related, for example, to AC Stark shifts and electric polarizabilities.
	We apply the method to the harmonic oscillator and the hydrogen atom, both driven by a time-dependent laser field.
	Supported by numerical simulations, we demonstrate the applicability to obtain relevant physical observables with high accuracy.
	The present approach offers a promising alternative for analytical studies of time-dependent multi-photon processes within the single-active-electron approximation and reduced-symmetry systems in the perturbative regime.
\end{abstract}

\maketitle
\newpage
\section*{Introduction}
In the non-relativistic regime, quantum dynamics is governed by the time-dependent Schr\"odinger equation (TDSE), which is a linear evolution problem.
In many situations, approximate yet accurate and physically relevant solutions can be calculated using time-dependent perturbation theory (TDPT), which represents a major tool for describing externally perturbed systems evolving under time-dependent interactions.
This approach is particularly effective when the coupling constant, which controls the strength of the interaction between the unperturbed system and the external perturbation, is \emph{sufficiently small}.
For instance, in attosecond physics, second-order TDPT underlies the theoretical description of the interferometric technique known as \emph{reconstruction of attosecond beating by interference of two-photon transitions} (RABBITT; see \cite{Paul}), as well as its multi-sideband extensions \cite{PhysRevA.54.721,Bharti}.
Within this framework, TDPT describes the photoelectron sideband oscillations, allowing one to extract the time delays that occur during photoionization.
In this context, the coupling constant is determined by the amplitudes of the infrared probe and the extreme-ultraviolet pump.

Standard TDPT is commonly formulated in terms of the \emph{Dyson series}, as presented in standard quantum mechanics textbooks; see, e.g. \cite{Sakurai,Schiff,Landau}.
Despite its wide success, the Dyson-series formulation becomes cumbersome at high orders due to explicit time ordering, the rapid growth of nested time integrals, and the eventual need to sum over intermediate states --- a conceptual difficulty recognized long ago~\cite{PhysRev.163.1343}.
Moreover, it offers limited scope for analytical treatments of the wave function, except for very specific perturbations (e.g. harmonic driving).
Furthermore, it typically involves infinite sums over the unperturbed spectrum of energies, whose accurate evaluation can be computationally expensive and whose convergence is not always guaranteed or is challenging to prove; see, e.g. \cite{Simon}.
A few alternative approaches to time-dependent problems are available in the literature, including \emph{stationary perturbation theory} \cite{Kutzelnigg} and \emph{Floquet perturbation theory} \cite{PhysRevA.56.4045,Rodriguez-Vega_2018}.

In contrast to TDPT, there are several formulations of its time-independent counterpart, among which \emph{Rayleigh--Schr\"odinger perturbation theory} \cite{Landau} is the textbook method par excellence.
A variety of alternative approaches have also been developed, including the methods of Dalgarno--Stewart~\cite{Dalgarno}, Fern\'andez and Castro~\cite{FC}, Swenson and Danforth~\cite{Swenson}, the \emph{moment method} \cite{PhysRevA.46.318}, the recently developed \emph{Green's function approach} \cite{Blankleider}, and \emph{logarithmic perturbation theory} (LPT)~\cite{Aharonov,Au}, also known as the \emph{non-linearization procedure}~\cite{Turbiner1984}, among others.
We refer the interested reader to~\cite{IntroPTQM} for a detailed overview and discussion of the scope of the various formulations, where their advantages and limitations are also described.
Among these formulations, LPT stands out due to the following features, which arise from considering the logarithm of the wave function as the central object: (i) it provides corrections of arbitrary order for the energy in integral form; and (ii) the unperturbed wave function serves as the main ingredient for constructing the series of the perturbed wave function.
Owing to these features, LPT has been extensively used to study a wide range of quantum systems, including anharmonic oscillators \cite{Dolgov1978}, screened Coulomb potentials \cite{ELETSKY1981235}, the Stark \cite{PhysRevA.22.1833} and Zeeman \cite{PhysRevA.103.032820} effects in hydrogen, as well as various radial potentials \cite{Aharonov}.
For a historical review of the method and its applications, we refer the reader to~\cite{Bandyopadhyay}.
To the best of the authors' knowledge, a time-dependent extension of LPT that preserves features (i) and (ii) has not been developed so far.

In this paper, we formulate the time-dependent extension of LPT with the aim of facilitating analytical results for time-evolved wave functions, which is often outside the scope of standard TDPT.
This method, here referred to as \emph{time-dependent logarithmic perturbation theory} (TDLPT), is designed to address the time-evolved phase (here defined as the logarithm of the wave function) and offers a natural way to compute dynamic energy shifts induced by an arbitrary external time-dependent potential.
The proposed approach is well-suited to situations in which the initial condition of the TDSE corresponds to a nondegenerate ground state of the unperturbed Hamiltonian --- a common physical scenario.
In addition to the formulation, we apply TDLPT to two physically relevant systems: (1)~the~one-dimensional harmonic oscillator; and (2) the hydrogen atom, both subjected to a linearly-polarized laser field.

In the case of the harmonic oscillator, we show that the exact solution of the TDSE can be fully recovered analytically within the TDLPT framework with only the first two correction terms, unlike in the standard formulation of TDPT.
Therefore, this application offers a proof of principle of the method.
In the case of the hydrogen atom, the analytical structure of the wave function at large distances can be established in the form of an asymptotic expansion.
In particular, we show that the correction terms to the wave function can be obtained by solving individual linear evolution problems, revealing that the time-evolved phase itself undergoes transitions that obey the standard selection rules.
We support this study with numerical solutions of the underlying evolution equations, which are then used to compute the AC energy shifts and the induced dipole moment.
For the latter, we benchmark against a numerical solution of the full TDSE.

The present paper is organized as follows.
We start by presenting a brief recapitulation of the time-independent version of LPT in section~\ref{Sec:LPT}.
Then, in section~\ref{Sec:Generalities}, we state the time-dependent problem under consideration.
In section~\ref{Sec:TDLPT}, we establish the theoretical framework of the TDLPT.
This includes the derivation of the equations that dictate the evolution of the correction terms and their expectation values.
Finally, it describes the connection to LPT and dynamic energy shifts.
We then apply our method to the harmonic oscillator and the hydrogen atom in section~\ref{Sec:Applications}.
Conclusions are presented in section~\ref{Sec:Conclusions}.

Supporting mathematical calculations and proofs are provided in the Appendices.
Atomic units ($\hbar=m_e=e=4\pi\varepsilon_0=1$) are used throughout the paper unless stated otherwise.

\section{Time-Independent Logarithmic Perturbation Theory}\label{Sec:LPT}
Before introducing the time-dependent extension, we briefly recall the basics of LPT for the time-independent Schr\"odinger equation\footnote{For a detailed description of LPT, we refer the reader to the seminal papers~\cite{Au,Aharonov,Turbiner1984}.
Here, we present only the minimal mathematical details needed below.
For simplicity, we assume $x\in\mathbb{R}^d$.}.
This will make the similarities with the formulation developed in the next section more transparent.

LPT was originally designed to find the ground state of the spectral problem
\begin{equation}\label{eq:SE}
	\left[-\frac{1}{2}\Delta+V_0(x)+\lambda V_{\rm int}(x)\right]\psi(x)
	=E\psi(x),
\end{equation}
where $V_0(x)$ and $V_{\rm int}(x)$ are time-independent potentials, and $\Delta$ denotes the Laplacian.

In contrast to standard perturbation theory, where the ansatz for the solution $\psi$ of \eqref{eq:SE} itself is expanded as a power series in the coupling constant $\lambda$, LPT uses the exponential representation for the (real-valued) wave function
\begin{equation}
	\psi(x)
	=\e^{-\phi(x)},
\end{equation}
together with the \emph{perturbation series}
\begin{equation}\label{eq:PT}
	\phi(x)
	=\sum_{n=0}^{\infty}\lambda^n\phi_n(x),\qquad
	E
	=\sum_{n=0}^{\infty}\lambda^n E_n.
\end{equation}
It is straightforward to show that the linear partial differential equation (PDE) defining $\phi_n$ and $E_n$ is given by
\begin{equation}\label{eq:LPT_divergence}
	\nabla\cdot\left[\psi_0(x)^2\nabla\phi_n(x)\right]
	=2\left(E_n-Q_n(x)\right)\psi_0(x)^2,\quad
	n\ge 1,
\end{equation}
where $\psi_0=\e^{-\phi_0}$ denotes the ground state wave function of the unperturbed problem ($\lambda=0$), and
\begin{equation}
	Q_1(x)
	=V_{\rm int}(x),\qquad
	Q_n(x)
	=-\frac{1}{2}\sum_{k=1}^{n-1}\nabla\phi_k(x)\cdot\nabla\phi_{n-k}(x),\quad
	n\ge 2.
\end{equation}
Integrating \eqref{eq:LPT_divergence} over $\mathbb{R}^d$, using the Gauss--Ostrogradsky Theorem, and assuming that the boundary term vanishes because of the decay of $\psi_0(x)$ at infinity, one obtains the integral expression
\begin{equation}\label{eq:LPT_energy_integral}
	E_n
	=\int_{\mathbb{R}^d}Q_n(x)\psi_0(x)^2\,\mathrm{d}x
	=\langle Q_n(x)\rangle_{\psi_0}.
\end{equation}
Thus, each energy correction $E_n$ is expressed as an expectation value of the corresponding static \emph{pseudopotential} $Q_n(x)$ with respect to the unperturbed ground-state density.

In one spatial dimension, \eqref{eq:LPT_divergence} can be solved explicitly, yielding $\phi_n(x)$ in integral form.
This makes the implementation of LPT straightforward and allows high orders in perturbation theory to be computed; see, for example, \cite{TurbinerDelValle2023}.
In contrast, multidimensional problems require sophisticated numerical methods, since exact solutions of \eqref{eq:LPT_divergence} are not available in general.
Exceptions occur in symmetry-reduced systems~\cite{Bandyopadhyay}, such as the hydrogen atom in a static multipole field~\cite{PhysRevA.22.328}, where the realization of the method becomes purely algebraic.

Although the roots of the method date back to the 1980s~\cite{Aharonov,Au,Turbiner1984}, and since then LPT has been widely applied to time-independent problems, to the best of our knowledge, no extension to explicitly time-dependent quantum dynamics has been developed so far.
In the following section, we formulate such an extension by applying a similar ansatz to the TDSE.
We will show that the resulting formulation closely parallels the time-independent case and describe the connection between the two.

\section{Time-Dependent Case: Generalities}\label{Sec:Generalities}
We consider the linear TDSE
\begin{equation}
	\begin{cases}\label{eq:main}
		\ii\,\partial_t\psi(x,t)
		=\hat{H}(t)\,\psi(x,t),\quad
		x\in\Omega,\quad
		t\ge 0,\\
		\psi(x,0)
		=\psi_0(x),
	\end{cases}
\end{equation}
where the domain $\Omega\subseteq\mathbb{R}^d$ is either bounded (thus, \eqref{eq:main} is equipped with suitable boundary conditions) or unbounded, and $\psi_0\in L^2(\Omega)$ is a given initial condition.
Moreover, we assume that $\hat{H}(t)$ is a time-dependent Hermitian operator, which can be written as the sum of a time-independent and a time-dependent component:
\begin{equation}\label{eq:partition}
	\hat{H}(t)
	=\hat{H}_0+\hat{H}_{\text{int}}(t).
\end{equation}
In practice, $\hat{H}_0$ is the free-system Hamiltonian taking the form
\begin{equation}
	\hat{H}_0
	=-\frac{1}{2}\Delta+V_0,\qquad 
	\Delta
	=\sum_{i=1}^{d}\partial_{x_i}^2,
\end{equation}
where the potential $V_0\colon{\mathbb{R}^d\to\mathbb{R}}$ is a real-valued function that acts as a multiplication operator.
As in many physical situations, we assume the initial condition to be the lowest normalized eigenfunction of $\hat{H}_0$, i.e.,
\begin{equation}\label{eq:zero-order}
	\hat{H}_0\,\psi_0
	=E_0\,\psi_0,\qquad
	\int_\Omega|\psi_0(x)|^2\,\mathrm{d}x
	=1,
\end{equation}
where $E_0$ denotes the corresponding lowest energy.
Therefore, $\psi_0$ represents the wave function of the ground state, which in particular is nodeless inside $\Omega$ and can always be assumed to be real-valued \cite{ReedSimonIV} and nondegenerate.
These assumptions on the initial condition are crucial for the present work.

Finally, the time-dependent component of the Hamiltonian is assumed to have the form
\begin{equation}
	\hat{H}_{\text{int}}(t)
	=\lambda\,V_{\text{int}}(x,t), 
\end{equation}
where $\lambda\in\mathbb{R}$ is \emph{small} and is referred to as the coupling constant.
Typically, the real-valued time-dependent potential $V_{\text{int}}$ arises from the interaction of the system with an external field.

In most situations of practical interest, the TDSE \eqref{eq:main} cannot be solved analytically, and one must therefore resort to numerical approximations.
Owing to the formidable challenges posed by this equation, the development of efficient and accurate numerical methods for approximating solutions of the TDSE remains an active area of research \cite{MR3562216,MR4858714}.
The development of combined analytical and numerical results is valuable since it sheds light on the underlying physical processes that purely numerical solutions of the TDSE may overlook.

Taking advantage of the smallness of the coupling constant, perturbation theory stands out as a well-suited methodology that may provide results of analytical nature.
Hence, in the spirit of standard LPT, here we will commence with an exponential \emph{ansatz} of the solution to \eqref{eq:main}.
As we will show, this consideration leads to various analytical results which have not been reported in the literature so far; for example, closed formulas of the corresponding correction terms and dynamic energy shifts, both in integral form.

\section{Time-Dependent Logarithmic Perturbation Theory}\label{Sec:TDLPT}
In a similar fashion to standard time-independent LPT \cite{Turbiner1984}, we consider the so-called \emph{exponential representation} of the wave function.
This means that we assume that the normalized solution of \eqref{eq:main} can be written formally\footnote{In the one-dimensional time-independent case, it is well-known that the exponential representation yields that zeros of the wave function appear as logarithmic singularities in the phase \cite{Aharonov,Turbiner1984}.
This feature is inherited to the time-dependent ansatz \eqref{eq:ExpRep}.
Similar exponential representations of the phase appear in quantum hydrodynamics or in the time-dependent WKB approximation and, to the best of our knowledge, they can only be justified by assuming their existence.} as
\begin{equation}\label{eq:ExpRep}
	\psi(x,t)
	=\e^{\Phi(x,t)}.
\end{equation}
By analogy with the time-independent counterpart, we refer to the complex function $\Phi(x,t)$ as the \emph{phase}.
To guarantee the square-integrability of the wave function in $\Omega=\mathbb{R}^d$, \eqref{eq:ExpRep}~enforces the growth condition
\begin{equation}
	\Re\,\Phi(x,t)\to-\infty
	\quad\text{as $|x|\to\infty$},\quad
	t\ge 0,
\end{equation}
which implies that the phase is unbounded.
By direct substitution of \eqref{eq:ExpRep} into \eqref{eq:main}, it can be shown that the phase obeys a nonlinear PDE, namely,
\begin{equation}\label{eq:non-linear}
	\ii\,\frac{\partial\Phi(x,t)}{\partial t}
	=-\frac{1}{2}\left[\Delta\Phi(x,t)+\nabla\Phi(x,t)^2\right]+V_0(x)+\lambda V_{\text{int}}(x,t).
\end{equation}
The initial phase $\Phi(x,0)$, and possible boundary conditions\footnote{In the case when $\Omega$ is bounded.}, of the previous equation are determined by \eqref{eq:main} and \eqref{eq:ExpRep}.
Thus, $\psi_0(x)=\e^{\Phi(x,0)}$, where $\psi_0$ is the ground state of $\hat H_0$ according to \eqref{eq:zero-order}.
In what follows, \eqref{eq:non-linear} will be solved perturbatively.

For $\lambda=0$ the solution of \eqref{eq:non-linear} is given by
\begin{equation}\label{eq:FreeEvolution}
	\Phi_0(x,t)
	=-\phi_0(x)-\ii\,E_0t,
\end{equation}	
where $\phi_0$ is such that
\begin{equation}\label{eq:defphi0}
	\psi_0(x)
	=\e^{-\phi_0(x)}.
\end{equation}
Indeed, upon inserting the exact solution \eqref{eq:FreeEvolution} into \eqref{eq:ExpRep}, the resulting wave function describes the well-known evolution $\psi(x,t)=\psi_0(x)\,\e^{-\ii\,t E_0}$ of a bound state in the absence of the time-dependent potential $V_{\rm int}(x,t)$.
The function $\phi_0$ satisfies a Riccati-like equation
\begin{equation}\label{eq:Riccati}
	\Delta\phi_0-(\nabla\phi_0)^2
	=2[E_0-V_0], 
\end{equation}
as can be shown by substituting \eqref{eq:FreeEvolution} into \eqref{eq:non-linear}.

\subsection{Perturbation Theory}
Following the standard scheme of perturbation theory, we construct a solution to \eqref{eq:non-linear} in the form of a power-series expansion\footnote{In this work, we do not study the convergence of this expansion, as such an analysis lies beyond the scope of the present paper and constitutes a challenging problem in its own right.
Therefore, the expansion should be regarded as formal at this stage.}
\begin{equation}\label{eq:main_series}
	\Phi(x,t)
	=\sum_{n=0}^\infty\lambda^n\,\Phi_n(x,t).
\end{equation}
In the following, we refer to the function $\Phi_n$ as the $n$th \emph{correction term}.
Note that truncating \eqref{eq:main_series} at order $N$ (i.e., retaining terms up to $\lambda^{N}$) generally breaks exact normalization of the wave function; however, the resulting deviation of the norm from unity appears only at order $\mathcal{O}(\lambda^{N+1})$.

A straightforward calculation shows that the $n$th correction term obeys the linear non-homogeneous PDE
\begin{equation}\label{eq:nth}
	\ii\frac{\partial\Phi_n(x,t)}{\partial t}
	=-\frac{1}{2}\left[\Delta\Phi_n(x,t)+2\nabla\Phi_0(x,t)\cdot\nabla\Phi_n(x,t)\right]+Q_n(x,t),\quad
	n\ge 1,
\end{equation}
where we define (this is indicated by the symbol $:=$ throughout the paper) the functions
\begin{equation}\label{eq:Qn}
	Q_1(x,t)
	:=V_{\text{int}}(x,t),\qquad
	Q_n(x,t)
	:=-\frac{1}{2}\sum_{k=1}^{n-1}\nabla\Phi_{n-k}(x,t)\cdot\nabla\Phi_{k}(x,t).
\end{equation}
We refer to $Q_{n\ge 1}$ as the $n$th pseudopotential.
Appropriate initial conditions must be imposed on \eqref{eq:nth} to ensure that the solution is consistent with \eqref{eq:FreeEvolution}.
In particular, these conditions are
\begin{equation}\label{eq:initial}
	\Phi_n(x,0)
	=0\quad
	\text{for all $n\ge 1$}.
\end{equation}
This choice is unique up to multiples of $2\pi\ii$, which would only introduce an irrelevant global phase to the solution of the Schr\"odinger equation.

The construction of the solution of \eqref{eq:nth}, one of the main goals of this work, is postponed to section~\ref{solutions}.
Next, we present related methods that share similarities with the approach presented here.

\subsection{Related Methods}
At first glance, the formulation presented above shares similarities with the ansatz 
\begin{equation}\label{eq:ansatz}
	\psi(x,t)
	=\sqrt{\rho(x,t)}\,\e^{\frac{\ii}{\hbar}S(x,t)}
\end{equation}
used in trajectory-based dynamics within the Madelung's hydrodynamic formulation of quantum mechanics \cite{Madelung,Wilhelm,Mera_2013}.
Additionally, if one equips \eqref{eq:ansatz} with the expansion
\begin{equation}\label{eq:action}
	S(x,t)
	=S_0(x,t)+\hbar\,S_1(x,t)+\ldots
\end{equation}
one arrives at the time-dependent ansatz employed in the WKB approximation \hbox{\cite{Wilhelm,Goldfarb,Lasser_Lubich_2020,leung2014fast}}.
However, there are important differences with the approach considered in the present paper.
(i)~In our formulation, the expansion parameter is the coupling constant $\lambda$, rather than the reduced Planck constant $\hbar$; see \eqref{eq:action}.
The latter is sometimes identified as the semiclassical parameter in the mathematical literature; see, e.g. \cite{Lasser_Lubich_2020,leung2014fast}.
(ii) In contrast to \eqref{eq:ansatz}, the exponential representation \eqref{eq:ExpRep} contains no prefactor (amplitude).
Furthermore, while $S(x,t)$ is restricted to be real, the phase $\Phi$ in \eqref{eq:ExpRep} typically consists of both non-zero real and imaginary parts.
Finally, we point out that there are also similarities with the formulation of TDPT presented by Chung \cite{PhysRev.163.1343}; however, in that case, the phase $S(x,t)$ is a function of time only and the exponential is accompanied by an amplitude expanded in powers of $\lambda$.

\subsection{The $n$th Order Correction\label{solutions}}
For each $n\ge 1$, equation~\eqref{eq:nth} together with the initial condition~\eqref{eq:initial} can be written in the form of an ordinary differential equation
\begin{equation}
	\ii\,\pa_tu(t)
	=\hat{\mathcal{L}}\,u(t)+f(t),\quad
	u(0)
	=0,
\end{equation}
where, in our specific setup, $\hat{\mathcal{L}}$ is a linear differential operator given by
\begin{equation}\label{eq:L}
	\hat{\mathcal{L}}
	:=-\frac{1}{2}\left(\Delta+2\nabla\Phi_0\cdot\nabla\right)
	\quad\text{and}\quad 
	f(t)
	:=Q_{n}(x,t).
\end{equation}
Using the relation $\nabla\Phi_0=-\nabla\phi_0$, which follows from \eqref{eq:FreeEvolution}, a straightforward calculation shows that, for any sufficiently smooth function $v$,
\begin{equation}
	\hat{\mathcal L}(\e^{\phi_0}v)
	=-\frac{1}{2}\e^{\phi_0}\left(\Delta+\Delta\phi_0-(\nabla\phi_0)^2\right)v.
\end{equation}
Equivalently,
\begin{equation}
	\e^{-\phi_0}\hat{\mathcal L}\e^{\phi_0}
	=-\frac{1}{2}\left(\Delta+\Delta\phi_0-(\nabla\phi_0)^2\right).
\end{equation}
Invoking the Riccati relation~\eqref{eq:Riccati}, it therefore follows that
\begin{equation}
	\e^{-\phi_0}\hat{\mathcal L}\e^{\phi_0}
	=\hat H_0-E_0,
\end{equation}
or, equivalently,
\begin{equation}\label{eq:formula_LR}
	\hat{\mathcal L}
	=\e^{\phi_0}\left(\hat H_0-E_0\right)\e^{-\phi_0}.
\end{equation}
Thus, $\hat{\mathcal L}$ is related to the shifted Hamiltonian $\hat H_0-E_0$ by a \emph{gauge rotation}.
Since similarity transformations are preserved under formal \emph{exponentiation}\footnote{From $L=QKQ^{-1}$, it follows $\e^{L}=Q\e^{K}Q^{-1}$.}, we obtain
\begin{equation}\label{eq:alternative}
	\e^{-\ii t\hat{\mathcal L}}
	:=\e^{\phi_0}\,\e^{-\ii t(\hat H_0-E_0)}\,\e^{-\phi_0}
	=\psi_0^{-1}\,\e^{-\ii t(\hat H_0-E_0)}\,\psi_0.
\end{equation}
As shown in Appendix~\ref{sec:representation}, $\hat{\mathcal{L}}$ is the generator of a strongly continuous unitary group $\e^{-\ii t\hat{\mathcal{L}}}$ on the weighted Hilbert space $L^2(\Omega,\psi_0^2\,\mathrm{d}x)$ and plays the role of the propagator.
Hence, a direct application of \emph{Duhamel's formula}\footnote{This formula is a consequence of Duhamel's principle; see \cite{Olver}.} \cite{MR2652783,kato1995} --- in the literature also known as the \emph{variation-of-constants formula} \cite{Pazy1983} --- yields the following closed integral representation for the correction terms; i.e., the solution of \eqref{eq:nth} for arbitrary $n\ge 1$:
\begin{align}
	\Phi_n(x,t)
	&=\e^{-\ii\,t\mathcal{L}}\,\Phi_n(x,0)-\ii\int_0^t\e^{-\ii(t-s)\,\hat{\mathcal{L}}}f(s)\,\mathrm{d}s\label{eq:arb_ini}\\
	&=-\ii\int_0^t\e^{\phi_0}\,\e^{-\ii (t-s)\,(\hat H_0-E_0)}\,\e^{-\phi_0}Q_n(x,s){\,\mathrm{d}s},\label{eq:formal_sol}
\end{align}
where in the last line the initial condition \eqref{eq:initial} was imposed on the general solution \eqref{eq:arb_ini}.
The availability of an integral representation for the $n$th correction allows the TDLPT wave function to be expressed as an exponential of explicitly computable terms.
In contrast, standard TDPT is formulated in terms of the Dyson series, with the relation between these two representations discussed in Appendix~\ref{app:connection}.

\subsection{Evolution Equation in Expectation-Value Form}
In this subsection, we derive the following evolution equation for the expectation values of the correction terms and the pseudopotentials $Q_n$:
\begin{equation}\label{eq:condition_expval}
	\ii\,\pa_t\langle\Phi_{n}(x,t)\rangle_{\psi_0}
	=\langle Q_n(x,t)\rangle_{\psi_0},\quad
	n\ge 1,
\end{equation}
where $\langle\cdot \rangle_{\psi_0}=\langle\psi_0|\cdot|\psi_0\rangle$ in Dirac notation with the ground state wave function $\psi_0$.
As we will show, this equation is independent of the imposed initial conditions for $\Phi_n$ and leads to the physical condition of the absence of current flux at the boundary of $\Omega$.
Furthermore, it is essential for connecting the TDPT with the time-independent case.

To show \eqref{eq:condition_expval}, we multiply both sides of the general solution \eqref{eq:arb_ini} by $\psi_0^2$ and integrate over the, possibly unbounded, domain $\Omega$.
This leads us to
\begin{equation}\label{eq:expectation}
	\int_\Omega\psi_0(x)^2\Phi_n(x,t)\,\mathrm{d}x
	=\int_\Omega\psi_0(x)^2\e^{-\ii\,t\hat{\mathcal{L}}}\Phi_n(x,0)\,\mathrm{d}x-\ii\int_0^t\int_\Omega\psi_0(x)^2\e^{-\ii(t-s)\,\hat{\mathcal{L}}}Q_n(x,s)\,\mathrm{d}x\,\mathrm{d}s.
\end{equation}
Moreover, using the gauge-rotated representation of $\e^{-\ii(t-s)\,\hat{\mathcal{L}}}$ together with \eqref{eq:defphi0} yields 
\begin{align}
	\int_\Omega\psi_0(x)^2\e^{-\ii(t-s)\,\hat{\mathcal{L}}}Q_n(x,s)\,\mathrm{d}x
	&=\int_\Omega\e^{-\phi_0}\,\e^{-\ii (t-s)\,(\hat H_0-E_0)}\,\e^{-\phi_0}Q_n(x,s)\,\mathrm{d}x\\
	&=\langle\psi_0,\hat{U}(t-s)\psi_0\,Q_n(x,s)\rangle,\label{eq:exp_Qn}
\end{align}
where $\hat{U}(t)=\e^{-\ii\,t(\hat{H}_0-E_0)}$ is the free-evolution operator on $L^2(\Omega)$.
Using the fact that $\hat{U}(t)$ is unitary and $\hat{U}(t)\psi_0=\e^{-\ii\,t0}\psi_0=\psi_0$ (recall that $\psi_0$ is the ground state of $\hat{H}_0$ and therefore $(\hat H_0-E_0)\psi_0=0$), the expectation value \eqref{eq:exp_Qn} can be rewritten as
\begin{equation}
	\langle\psi_0,\hat{U}(t-s)\psi_0\,Q_n(x,s)\rangle
	=\langle\hat{U}^{\dagger}(t-s)\psi_0,\psi_0\,Q_n(x,s)\rangle
	=\langle Q_n(x,s)\rangle_{\psi_0}.
\end{equation}
Thus, the explicit dependence on $t$ drops out of this term.
Using a similar procedure, one can show that
\begin{equation}
	\int_\Omega\psi_0(x)^2\e^{-\ii\,t\mathcal{L}}\Phi_n(x,0)\,\mathrm{d}x
	=\langle\Phi_n(x,0)\rangle_{\psi_0}.
\end{equation}
This implies that \eqref{eq:expectation} can be presented equivalently as
\begin{equation}
	\langle\Phi_n(x,t)\rangle_{\psi_0}
	=\langle\Phi_n(x,0)\rangle_{\psi_0}-\,\ii\int_0^t\langle Q_n(x,s)\rangle_{\psi_0}\,\mathrm{d}s.
\end{equation}
After taking the time derivative, the stated result \eqref{eq:condition_expval} emerges.
This result has direct implications on the particle current at the boundary $\partial\Omega$.
Indeed, multiplying \eqref{eq:nth} by $\psi_0^2$ and integrating over $\Omega$, we obtain the following relation:
\begin{equation}
	\ii\,\pa_t\langle\Phi_{n}(x,t)\rangle_{\psi_0}
	=-\frac{1}{2}\int_{\Omega}\nabla\cdot\left[\psi_0(x)^2\nabla\Phi_n(x,t)\right]\mathrm{d}x+\langle Q_n(x,t)\rangle_{\psi_0}.
\end{equation}
Consequently, \eqref{eq:condition_expval} enforces
\begin{equation}
	\int_{\Omega}\nabla\cdot\left[\psi_0(x)^2\nabla\Phi_n(x,t)\right]\mathrm{d}x
	=0.
\end{equation}
In the case where $\Omega$ is a bounded domain, one can apply the Gauss--Ostrogradsky Theorem to the latter equation; therefore,
\begin{equation}
	\int_{\partial\Omega}\psi_0(x)^2\,\nabla\Phi_n(x,t)\cdot\mathbf{n}(x)\,\mathrm{d}S
	=0,
\end{equation}
where $\mathbf{n}(x)$ denotes the outward unit normal on $\partial\Omega$.
This implies that, for all times, there is no current flux through the boundary \cite{Turbiner1984}.

\subsection{Recovering Time-Independent LPT}
Standard (time-independent) LPT can be recovered in the present (time-dependent) approach from the evolution equation in expectation-value form \eqref{eq:condition_expval}.
To do so, let us assume that the potential $V_{\rm int}$ is time-independent for the time being.
In this case, the dynamics of a stationary state of the TDSE, 
\begin{equation}\label{eq:mainTISE}
	\ii\,\partial_t\psi(x,t)
	=[\hat{H}_0+\lambda\,V_{\rm int}(x)]\,\psi(x,t),
\end{equation}
evolves according to
\begin{equation}
	\psi(x,t)
	=\e^{-\ii\,Et}\,\Psi(x),
\end{equation}
where the energy $E$ and the initial condition $\Psi$ are defined through
\begin{equation}\label{eq:TISE}
	[\hat{H}_0+\lambda\,V_{\rm int}]\Psi
	=E\,\Psi,
\end{equation}
i.e., the initial condition is given by the eigenfunction associated to $E$.
For simplicity of the exposition, the dependence on $\lambda$ is omitted in both $E$ and $\Psi$.
We point out that, when comparing \eqref{eq:main} to \eqref{eq:mainTISE}, the nature of the problem has changed.
The former describes time evolution starting from the unperturbed state, whereas the latter describes the stationary ground state of the already perturbed Hamiltonian.
Thus, when $\lambda=0$, the solution of the equation \eqref{eq:TISE} coincides with that given in \eqref{eq:defphi0}.

For $\lambda\ne 0$, one considers the following expansions:
\begin{equation}
	\Psi(x)
	=\e^{-\sum_{n=0}^\infty\lambda^n\phi_n(x)},\qquad
	E
	=\sum_{n=0}^{\infty}\lambda^nE_n,
\end{equation}
in agreement with \eqref{eq:PT}.
Therefore, it follows from \eqref{eq:ExpRep} and \eqref{eq:main_series} that the correction terms now have the form
\begin{equation}\label{eq:TI}
	\Phi_n(x,t)
	=-\phi_n(x)-\ii\,E_n t,\quad 
	n\ge 0,
\end{equation}
see \eqref{eq:initial}.
Substituting~\eqref{eq:TI} into~\eqref{eq:condition_expval} yields the well-established integral expression \cite{Au,Turbiner1984,Imbo,Rogers,PRIVMAN1981326,DOLGOV1978403,Coutinho,GALIAUTDINOV201872,Vikhnina} for the energy correction $E_n$ in terms of the $n$th pseudopotential presented in \eqref{eq:LPT_energy_integral} --- one of the hallmarks of time-independent LPT.

\subsection{Integral Form of Dynamic Energy Shifts}
Upon restoring physical constants, $\langle Q_n(x,t)\rangle_{\psi_0}$ carries units of energy; see~\eqref{eq:Qn}.
Therefore, as suggested by equation~\eqref{eq:LPT_energy_integral}, the following definition
\begin{equation}\label{eq:instantaneous_energy}
	E_n(t)
	:=\langle Q_n(x,t)\rangle_{\psi_0},
\end{equation}
may be interpreted as an \emph{instantaneous energy shift}, which reduces to the standard energy correction \eqref{eq:LPT_energy_integral} in the time-independent case. 
By taking its time average, we may further define the corresponding $n$th order (possibly complex) \emph{dynamic energy shift}\footnote{For $n=1$, this formula coincides with the one reported in the literature \cite{RevModPhys.44.602,Rodriguez-Vega_2018}.}
\begin{equation}\label{eq:average_energy_correction}
	\overline{E}_n
	:=\frac{1}{T}\int_{t_0-\frac{T}{2}}^{t_0+\frac{T}{2}}\langle Q_n(x,t)\rangle_{\psi_0}\,\mathrm{d}t.
\end{equation}
Here, $T>0$ denotes the length of the averaging time window, and $t_0$ denotes some given reference time of interest.
In this way, one can define dynamic energy shifts for any time-dependent perturbation.
Note that the dependence of $\overline{E}_n$ on both $T$ and $t_0$ is omitted in our notation.
Expression~\eqref{eq:average_energy_correction} thus provides a prescription for calculating, for example, dynamic (AC) Stark shifts induced by an arbitrary electric field\footnote{Not necessarily arising from a harmonic interaction, as typically assumed in the literature \cite{Sakurai}.}, as well as the closely related dipole polarizabilities.
This will be illustrated by the specific applications discussed in section~\ref{Sec:Applications}.

\section{Applications}
\label{Sec:Applications}
In this section, we explore two examples where the previously developed formalism of TDLPT is applied.
Given their physical relevance, we consider first the harmonic oscillator subjected to an external time-dependent electric field.
This example provides a proof of principle, since the time evolution of the system can be determined exactly \cite{Kerner}.
We then apply the formalism to study the dynamics of the hydrogen atom under the influence of a laser pulse, focusing on the first-order correction term, the dynamic energy shifts, the induced dipole moment, and the asymptotic expansion of the wave function via the formal solution \eqref{eq:formal_sol}.

\subsection{Harmonic Oscillator in an External Electric Field}
Consider a one-dimensional harmonic oscillator, initially in its ground state, perturbed by a time-dependent electric field.
Thus, we take the Hamiltonian
\begin{equation}
	\hat{H_0}
	=-\frac{1}{2}\partial_{x}^2+\frac{1}{2}x^2,\quad
	-\infty<x<\infty,
\end{equation}	
and the time-dependent external potential
\begin{equation}
	V_{\text{int}}(x,t)
	=x\,g(t),
\end{equation}	
where $g(t)$ describes the temporal profile of the external electric field.
Therefore, the amplitude of the electric field plays the role of the coupling constant.
Since the initial condition is the ground state of $\hat{H}_0$, we have
\begin{equation}
	\psi_0(x)
	=\e^{-\phi_0(x)},\qquad
	\phi_0(x)
	=\frac{x^2}{2}+\frac{1}{4}\ln\pi.
\end{equation}
According to \eqref{eq:FreeEvolution} this implies that
\begin{equation}
	\Phi_0(x,t)
	=-\frac{x^2}{2}-\frac{1}{4}\ln\pi-\ii\,E_0 t,\qquad
	E_0
	=\frac{1}{2}.
\end{equation}
As a consequence, \eqref{eq:L} reads
\begin{equation}
	\hat{\mathcal{L}}
	=-\frac{1}{2}\partial_x^2+x\,\pa_x,
\end{equation}
which is, up to a global and irrelevant constant factor, the Hermite operator, i.e., the standard differential operator appearing in Hermite's differential equation.
Following \eqref{eq:arb_ini}, the first-order correction term reads
\begin{equation}
	\Phi_1(x,t)
	=-\ii\int_0^t\e^{-\ii(t-s)\hat{\mathcal{L}}}\left[xg(s)\right]\,\mathrm{d}s
	=-\ii\int_0^tg(s)\,\e^{-\ii(t-s)\hat{\mathcal{L}}}x\,\mathrm{d}s.
\end{equation}	
The action of $\e^{-\ii(t-s)\hat{\mathcal{L}}}$ on the function $x$ can be carried out trivially by noticing that the function $x$ is an eigenfunction of $\hat{\mathcal{L}}$ with unit eigenvalue, i.e.,
\begin{equation}
	\hat{\mathcal{L}}\,x
	=x.
\end{equation}
An analogous argument applies when the explicit gauge-rotated representation \eqref{eq:alternative} is used.
Consequently,
\begin{equation}\label{eq:Phi_1}	
	\Phi_1(x,t)
	=-\ii\,x\int_0^tg(s)\,\e^{-\ii(t-s)}\,\mathrm{d}s,
\end{equation}	
which, according to \eqref{eq:Qn}, yields $Q_2=-\tfrac{1}{2}\left[\partial_x\Phi_1\right]^2$.
Hence, the next order correction reads
\begin{equation}\label{eq:Phi_2}
	\Phi_2(x,t)
	=\frac{\ii}{2}\int_0^t\e^{\ii(s-t)\hat{\mathcal{L}}}\left[\partial_x\Phi_1(x,s)\right]^2\,\mathrm{d}s,
\end{equation}	
which, as seen from \eqref{eq:Phi_1}, has no $x$-dependence.
Thus, we conclude that
\begin{equation}
	\Phi_2(x,t)
	=-\frac{\ii}{2}\int_0^t\int_0^s\int_0^s\e^{-\ii(2s-\tau-u)}g(\tau)g(u)\,\mathrm{d}\tau\,\mathrm{d}u\,\mathrm{d}s.
\end{equation}
A straightforward computation shows that the relations \eqref{eq:condition_expval} are satisfied, confirming the consistency of the solution constructed above.
Note that $E_1(t)=\langle Q_1(x,t)\rangle_{\psi_0}=0$, which confirms that no dynamic linear Stark shift is induced in the system.

Since $\Phi_2(x,t)$ does not depend on $x$, higher-order corrections vanish, i.e., $\Phi_{n}\equiv 0$ for all $n>2$.
Therefore, the series in \eqref{eq:main_series} is naturally truncated after only three terms.
The resulting expression for the wave function, $\e^{\Phi_0+\lambda\Phi_1+\lambda\Phi_2}$, thus provides the exact solution of the TDSE.
Let us point out that standard perturbation theory based on the Dyson series, see \eqref{eq:Dyson}, contains infinitely many terms in the expansion of the wave function, whereas TDLPT requires only three.
Note that the Dyson series can be easily recovered from the expansion of TDLPT by expanding $\e^{\Phi_0+\lambda\Phi_1+\lambda^2\Phi_2}$ in powers of~$\lambda$.

We now consider the case in which the electric field is switched on (and off) adiabatically at infinity and corresponds to a monochromatic laser field of frequency $\omega$.
The corresponding temporal profile of the field is given by
\begin{equation}
	g(t)
	=\e^{-\epsilon |t|}\cos(\omega t),\quad
	-\infty<t<\infty,
\end{equation}
for which the limit $\epsilon\to0^{+}$ must be taken at the end of the calculation to reflect the adiabatic consideration; see, e.g.~\cite{Joachain} for details.
Therefore, the first-order correction takes the form
\begin{equation}\label{eq:adiabatic}
	\Phi_1(x,t)
	=-\ii\,x\lim_{\epsilon\to 0^{+}}\int_{-\infty}^t\!\!\!\!\e^{-\epsilon|s|}\cos(\omega s)\,\e^{-\ii(t-s)}\,\mathrm{d}s.
\end{equation}
Note the consistent change\footnote{If a vanishing initial condition for the $n$th correction term is imposed at $t=t_0$, then $$\Phi_n(x,t)=-\ii\int_{t_0}^t\! \e^{-\ii (t-s)\hat{\mathcal{L}}}\,Q_n(x,s)\,\mathrm{d}s.\\[-5mm]$$} in the lower limit of integration, see \eqref{eq:Phi_1}.
In this case, the dynamic energy shifts, obtained after taking the time-average over one optical cycle of the laser field, can be calculated analytically; as seen in Appendix~\ref{app:Stark}.
From the explicit expressions~\eqref{eq:Phi_1} and~\eqref{eq:Phi_2}, together with equation~\eqref{eq:average_energy_correction}, it follows that only $\overline{E}_2$ is non-vanishing and results in
\begin{equation}
	\overline{E}_2
	=\frac{1}{4(\omega^{2}-1)}.
\end{equation}
This result\footnote{The same outcome is obtained if the laser is switched on \emph{nonadiabatically} at $t=0$ and the limit $T\to\infty$ is taken in \eqref{eq:average_energy_correction}.} coincides with the quadratic (AC) Stark shift induced on the $j$th eigenstate ($j=0$ in the present case) obtained by means of the standard expression developed for a monochromatic laser field, namely
\begin{equation}\label{eq:standard}
	\overline{E}_2(j)
	=-\frac{1}{2}\sum_{k\ne j}\frac{\bigl|\langle\psi_k|{\boldsymbol{\epsilon}}\cdot\hat{\mathbf{D}}|\psi_j\rangle\bigr|^{2}\,(E_k-E_j)}{(E_k-E_j)^2-\omega^2},
\end{equation}
where $\boldsymbol{\epsilon}$ denotes the polarization vector and $\hat{\mathbf{D}}$ the dipole operator (see, e.g. \cite{Joachain} for details).

\subsection{Hydrogen Atom Subjected to a Linearly Polarized Laser Field}\label{sec:Hydrogen}
Within the dipole approximation and assuming an infinitely massive proton, a hydrogen atom interacting with a laser pulse of temporal profile $g(t)$ is described by the nonrelativistic Hamiltonian
\begin{equation}
	\hat{H}(t)
	=-\frac{1}{2}\Delta-\frac{1}{r}-\lambda\,g(t)z,\qquad
	r
	=\sqrt{x^2+y^2+z^2}.
\end{equation}
A convenient decomposition of this Hamiltonian, according to \eqref{eq:partition}, is given by
\begin{equation}\label{eq:choice}
	\hat{H}_0
	=-\frac{1}{2}\Delta-\frac{1}{r},\qquad
	V_\text{int}(t)
	=-g(t)\,z.
\end{equation}
Additionally, we assume that the atom is initially in its ground state.
Therefore,
\begin{equation}
	\psi_0
	=\frac{1}{\sqrt{\pi}}\,\e^{-r},\quad\text{and}\quad
	\Phi_0(r,t)
	=-r-\frac{1}{2}\log\pi-\ii E_0t,
\end{equation}
with $E_0=-1/2$.

For convenience in the following calculations, we use the set of nonorthogonal \emph{hybrid} coordinates $\{r,\varphi,z\}$, see Appendix~\ref{app:coordinates} for details.
This choice of coordinates is motivated by the time-independent LPT applied to the Stark effect in the hydrogen atom ground state induced by a constant electric field.
In this case, the construction of the perturbation series becomes an algebraic procedure where the phase corrections \eqref{eq:TI} acquire polynomial form in the variables\footnote{The ground state is independent of $\varphi$, as dictated by symmetry.} $(r,z)$; see \cite{Turbiner1983Hydrogen}.
This suggests that the present choice of coordinates may provide a natural framework for extending the polynomial structure to the time-dependent case.

According to \eqref{eq:Laplacian} and \eqref{eq:grad_dot_grad}, the generator $\hat{\mathcal{L}}$ in \eqref{eq:L} can be expressed in the hybrid coordinates as
\begin{equation}
	\hat{\mathcal{L}}
	=-\frac{1}{2}\left(\pa_r^2+\frac{2}{r}\pa_r+\frac{2z}{r}\pa_{rz}+\pa_z^2+\frac{1}{r^2-z^2}\pa_{\varphi}^2\right)+\partial_{r}+\frac{z}{r}\,\partial_{z}.
\end{equation}	
In practice, derivatives with respect to $\varphi$ are irrelevant for the phase evolution due to the cylindrical symmetry of the system.
For this reason, we can neglect the $\pa_\varphi^2$ term from the above expression.

\subsubsection{First-Order Correction}
In contrast to the harmonic oscillator application, the action of $\e^{-\ii t\hat{\mathcal{L}}}$ on $Q_1=V_{\text{int}}$ cannot be evaluated directly.
However, we can resort to the PDE \eqref{eq:nth}, rewritten in hybrid coordinates, which defines the corrections and solve it by either numerical or analytical means.
In particular, the first-order correction satisfies
\begin{gather}\label{eq:H_first}
	\ii\,\pa_t\Phi_1(r,z,t)
	=\left[-\frac{1}{2}\left(\partial^2_{r}+\frac{2}{r}\,\partial_{r}+\frac{2z}{r}\,\partial_{rz}+\partial^2_{z}\right)+\partial_{r}+\frac{z}{r}\,\partial_{z}\right]\Phi_1(r,z,t)-g(t)z,
\end{gather}
with initial condition $\Phi_1(r,z,0)=0$.
We next consider the following factorization 
\begin{equation}\label{eq:factorization}
	\Phi_{1}(r,z,t)
	=\frac{z\,\e^{r}}{r^{2}}\,\varPhi_{1,1}(r,t),
\end{equation}
where $\varPhi_{1,1}$ is a certain function.
A straightforward calculation shows that this ansatz removes the first-order derivative terms in $r$, together with the explicit $z$-dependence of the equation \eqref{eq:H_first}, and leads to
\begin{equation}\label{eq:first}
	\ii\,\partial_{t}\varPhi_{1,1}(r,t)
	=-\frac{1}{2}\,\partial_{r}^2\varPhi_{1,1}(r,t)+\left(-\frac{1}{r}+\frac{1}{r^{2}}+\frac{1}{2}\right)\varPhi_{1,1}(r,t)-r^{2}\e^{-r}g(t),
\end{equation}
which we equip with $\varPhi_{1,1}(r,0)=0$.
Remarkably, upon neglecting the inhomogeneous term, \eqref{eq:first} resembles the evolution equation of a hydrogen $p$ state up to an overall energy shift.
Therefore, it may be regarded as a TDSE with a source term.

As shown below, the appropriate (asymptotic) boundary conditions for this equation can be justified by asymptotic analysis, and are given by
\begin{equation}\label{eq:boundary}
	\varPhi_{1,1}(0,t)
	=\varPhi_{1,1}(\infty,t)
	=0.
\end{equation}
For $r\to 0$, we use the following ansatz for the leading-order term 
\begin{equation}\label{eq:small_r}
	\varPhi_{1,1}(r,t)
	\sim a(t)\,r^\alpha,
\end{equation}
where both the function $a(t)$ and the exponent $\alpha$ are to be determined.
Substituting \eqref{eq:small_r} into \eqref{eq:first} and imposing asymptotic matching and the normalizability of \eqref{eq:ExpRep} up to $\mathcal{O}(\lambda^2)$ yields $\alpha=2$.
Hence, $\varPhi_{1,1}(0,t)=0$.
For $r\to\infty$, we consider the solution of \eqref{eq:first}, namely
\begin{equation}
	\varPhi_{1,1}(r,t)
	=\ii\int_0^t g(s)\,\e^{-\ii(t-s)\hat{\mathcal{L}}_r}\!\left[r^2\e^{-r}\right]\mathrm{d}s
	\quad\text{with}\quad
	\hat{\mathcal{L}}_r
	:=-\frac{1}{2}\partial_r^2+\left(-\frac{1}{r}+\frac{1}{r^{2}}+\frac{1}{2}\right),
\end{equation}
and analyze the formal action of the gauge-rotated unitary group generated by $\hat{\mathcal{L}}_r$.
Using the first two terms in the exponential series for $\e^{-\ii(t-s)\hat{\mathcal{L}}_r}$, we can establish
\begin{equation}
	\varPhi_{1,1}(r,t)
	=\ii\,r^2\e^{-r}\int_0^t\!\left[1+\frac{\ii(s-t)}{r}+\mathcal{O}(r^{-2})\right]g(s)\mathrm{d}s,
\end{equation}
which implies $\varPhi_{1,1}(\infty,t)=0$.
Moreover, the high-order extension of the large-$r$ asymptotic expansion
delivers a more accurate\footnote{This follows from a resummation of the leading and subleading contributions generated by repeated application of $\e^{-\ii(t-s)\hat{\mathcal{L}}}$ to the unity.} representation of $\varPhi_{1,1}$, namely
\begin{equation}
	\varPhi_{1,1}(r,t)
	=\ii\,r^2\e^{-r}\!\int_0^t\left[h(\xi)+\mathcal{O}(r^{-8})\right]g(s)\,\mathrm{d}s,
\end{equation}
where we have introduced
\begin{equation}
	h(\xi)
	:=\xi+\frac{\xi-1-\log\xi}{r}+\frac{\xi-1}{r^2}\left[\frac{(1+\xi)(\xi-1)}{2\xi^2}-\frac{\log\xi}{\xi}\right],
	\quad\xi
	:=1+\ii(s-t)/r.
\end{equation}
This, in turn, reveals the large-$r$ behavior of the wave function to first order in $\lambda$: 
\begin{equation}\label{eq:asymptotics_WF}
	\psi(r,t)
	=\frac{1}{\sqrt{\pi}}\e^{-r+\frac{\ii}{2}t}\left\{1+\ii\,\lambda\,z\int_0^t\left[h(\xi)+\mathcal{O}(r^{-8})\right]g(s)\,\mathrm{d}s+\mathcal{O}(\lambda^2)\right\}. 
\end{equation}
We emphasize that this result holds for an arbitrary temporal profile of the pulse.
To the best of the authors' knowledge, this expression, which makes the structure of the wave function explicit, has not been reported in the literature and illustrates the capability of TDLPT to provide analytical results.
While \eqref{eq:asymptotics_WF} can be extended to arbitrarily high order in powers of $r^{-1}$, the calculations quickly become cumbersome and are outside the scope of the present work.

After establishing results of analytical nature, we now solve equation~\eqref{eq:first} numerically.
In particular, we employed a Crank--Nicolson integrator with second-order central differences for the spatial discretization, achieving high accuracy and efficiency.
In this numerical implementation, the radial coordinate is discretized with a uniform grid spacing $dr=0.1$.
Moreover, the asymptotic boundary conditions \eqref{eq:boundary} are well approximated by enforcing $\varPhi_{1,1}(R_{\min},t)=\varPhi_{1,1}(R_{\max},t)=0$, with $R_{\min}=10^{-6}$ and $R_{\max}=40$.
In turn, the time evolution is performed until the pulse is over using a timestep $dt=0.001$. 

In our simulations, we adopt a laser pulse with the widely used $\sin^2$ temporal envelope, namely
\begin{equation}\label{eq:laser}
	g(t)
	=\sin^2\left(\frac{\omega t}{2N}\right)\cos(\omega t).
\end{equation}
Here, $\omega>0$ is the central frequency of the driving field, and $N\ge 1$ denotes the number of optical cycles in the pulse.
The electric field is taken to be nonzero only within the time interval $0\le t\le 2N\pi/\omega$, which implies that the final simulation time is $T_f=2N\pi/\omega$.
With \eqref{eq:laser}, the coupling constant corresponds to the amplitude of the electric field~$\mathcal{E}_0$,
\begin{equation}
	\lambda
	=\mathcal{E}_0
	=\sqrt{\frac{2I_0}{\varepsilon_0 c}},
\end{equation}
where $I_0$ is the peak laser intensity, $\varepsilon_0$ is the vacuum permittivity, and $c$ is the speed of light.
Specifically, the following set of parameters in atomic units was used in the simulations: $\omega=0.056$, $I_0=9\times10^{-4}$, and a representative number of optical cycles ranging from $N=5$ to $N=50$.
These parameters correspond to experimentally accessible conditions for a near-infrared Ti:sapphire laser with central wavelength $\simeq 800\,\mathrm{nm}$ and peak intensity $I_0=3.2\times10^{13}\,\mathrm{W/cm^2}$, yielding an electric-field amplitude $\mathcal{E}_0=0.03$.
We note, however, that the amplitude does not play any role while solving \eqref{eq:first}, but its \emph{small} value ensures the perturbative treatment.
\begin{table}[]
	\caption{Dynamic energy shift as a function of $N$ for different time windows: (i) single optical cycle; (ii) duration of the pulse, see text for details.
		For case (i) the dipole polarizability $\alpha(\omega)$ is reported in the third column; see \eqref{eq:polarizability}.
		The last row $(N=\infty)$ corresponds to the infinite-cycle limit calculated through \eqref{eq:standard}.
		Results are presented in atomic units and rounded to displayed decimal digits.}
	{\setlength{\tabcolsep}{0.4cm}		
	\begin{tabular}{ccc|c}
		\hline
		\hline
		$N$ & $\overline{E}_2(\text{cycle})$ & $\alpha(\omega)$&$\overline{E}_2(\text{pulse})$\\
		\hline
		5 & $-1.067$ & $4.267$ & $-0.430$\\
		10 & $-1.126$ & $4.504$ & $-0.430$\\
		15 & $-1.138$ & $4.550$ & $-0.430$\\
		20 & $-1.142$ & $4.567$ & $-0.430$\\
		30 & $-1.144$ & $4.579$ & $-0.430$\\
		50 & $-1.146$ & $4.583$ & $-0.430$\\
		$\infty$ & $-1.146$ & $4.585$ & $-$\\
		\hline
		\hline
	\end{tabular}
	}
	\label{tab:shifts_and_polarizabilities}
\end{table}

Once the function $\varPhi_{1,1}$ is determined numerically, the dynamic energy shift is evaluated using \eqref{eq:average_energy_correction} by performing the average (i) over a time window of one optical cycle, \hbox{$T=2\pi/\omega$}, centered at $t_0=N\pi/\omega$, where the pulse reaches its maximum, and (ii) over the full duration of the pulse, $T=2t_0=2N\pi/\omega$.
For both cases, results are shown in Table~\ref{tab:shifts_and_polarizabilities} for representative numbers of cycles.
As expected, the dynamic energy shift is real in both cases.
Moreover, in case (i) the energy shifts converge, as the number of cycles increases, to the value corresponding to an infinitely long monochromatic pulse\footnote{As $N$ increases, the envelope $\sin^2$ tends to the constant value $1$ near the pulse maximum, so the field \eqref{eq:laser} is well approximated by the monochromatic pulse $\cos(\omega t)$.}.
In Table~\ref{tab:shifts_and_polarizabilities}, we also report the \emph{dynamic polarizability} $\alpha(\omega)$ defined by the relation
\begin{equation}\label{eq:polarizability}
	\overline{E}_2
	=-\frac{1}{2}\alpha(\omega)\,\overline{g(t)^2}.
\end{equation}
As exhibited by the third column, we observe convergence of the polarizability to the monochromatic case once $N\to\infty$.
For case (ii), the dynamic energy shift $\overline{E}_2$ is found to be independent of $N$, as seen from the fourth column --- a rather unexpected result.
We are not aware of a similar observation in the literature.
 
\begin{figure}[]
	\centering
	\includegraphics[width=\linewidth]{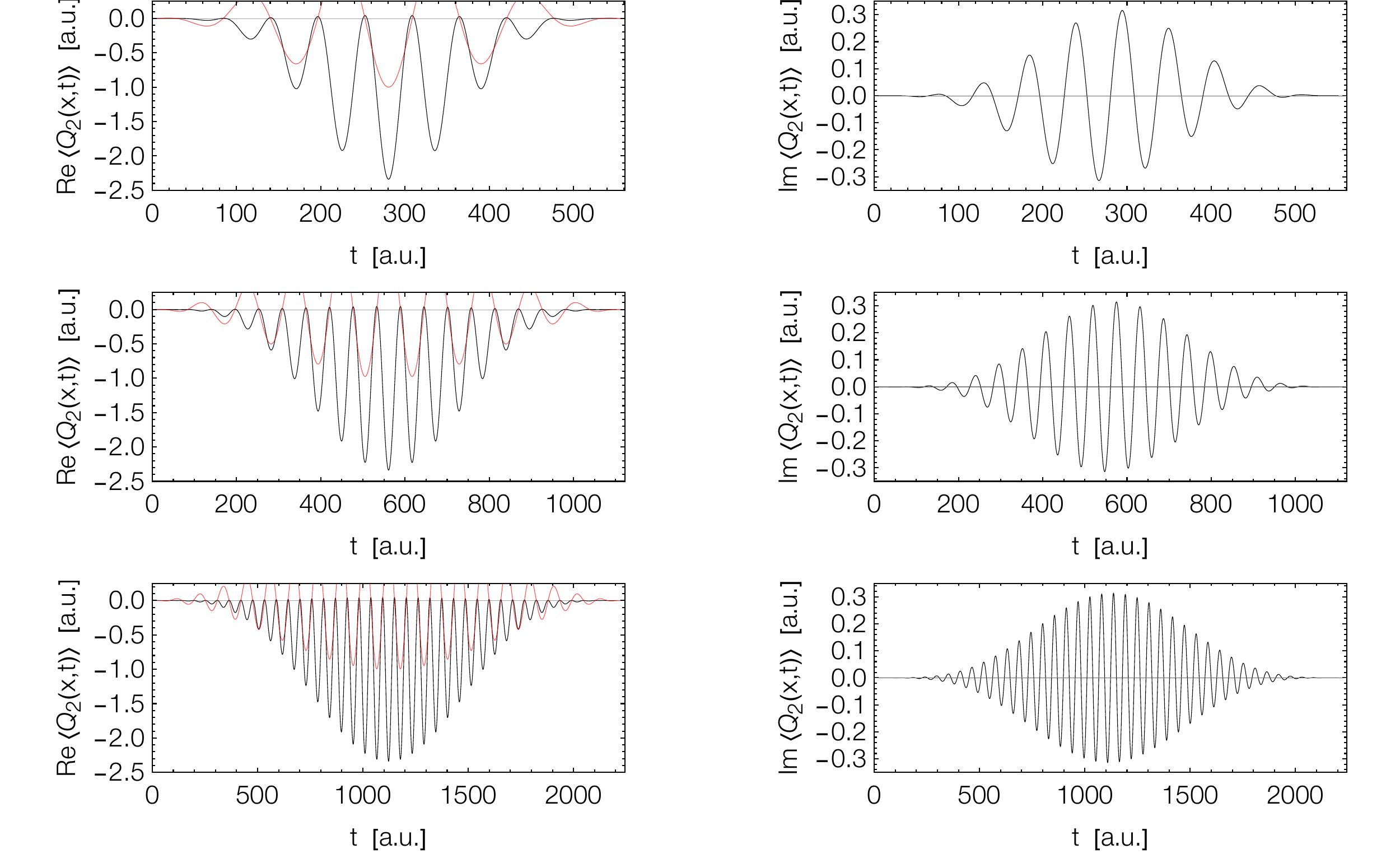}
	\caption{Real and imaginary parts of the instantaneous energy shift \eqref{eq:instantaneous_energy} as functions of time for $N=5\text{\,(top)},10\,\text{(center)},20\,(\text{bottom})$ optical cycles.
		The temporal profile of the field is shown in red (left panels) with $\omega=0.056$~a.u. ($\simeq800$~nm); see \eqref{eq:laser}.}
	\label{fig:Instantaneous_Energy}
\end{figure}

In addition, we show in figure~\ref{fig:Instantaneous_Energy} the real and imaginary parts of the instantaneous energy shifts \eqref{eq:instantaneous_energy} as a function of time for $N=5,10,20$.
As we can see, the energy shifts are driven by the electric field, the real part oscillates in phase with it, and the atomic response to the field is more pronounced near the peak of the pulse.

\begin{figure}[b]
	\centering
	\includegraphics[scale=0.4]{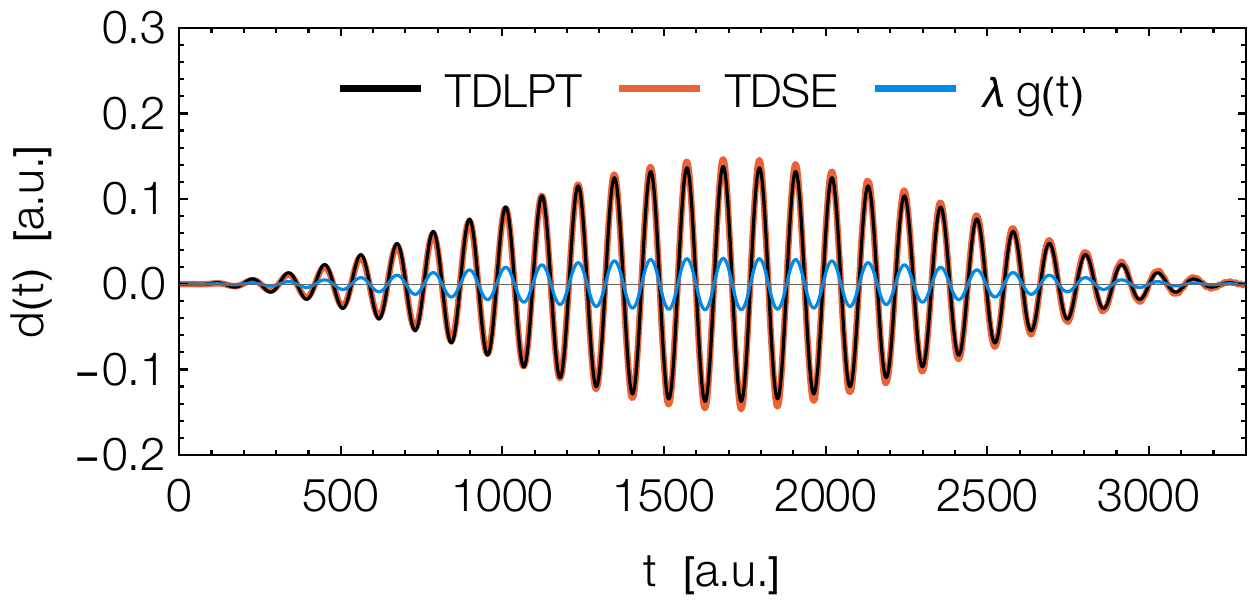}
	\caption{Time-dependent induced dipole moment calculated using first-order TDLPT and the full TDSE solution (black and red curves, respectively).
		The corresponding laser field is shown in blue.
		Parameters of Eq.~\eqref{eq:laser}: $N=30$ optical cycles, $\omega=0.056$~a.u. (corresponding to $\simeq 800$~nm), and $\mathcal{E}_0=0.03$~a.u.}
	\label{fig:dipole}
\end{figure}

Further insight is provided by the time-dependent induced dipole moment $d(t)$ along the $z$~direction, which is given by
\begin{equation}
	d(t)
	=-\int_{\mathbb{R}^3} z\,|\psi(x,y,z,t)|^2\,\mathrm{d}x\,\mathrm{d}y\,\mathrm{d}z
	=-\frac{8}{3}\,\lambda\,\Re\int_0^\infty\!\!\!\varPhi_{1,1}(r,t)\e^{-r}r^2\,\mathrm{d}r+\mathcal{O}(\lambda^2).
\end{equation}
Working at first order in perturbation theory, terms of order $\mathcal{O}(\lambda^2)$ and higher are neglected in our numerical calculation.
To benchmark the TDLPT results, we constructed a highly accurate TDSE reference solution and computed the dipole moment using the well-established computational code described in \cite{atoms12070034}.
Figure~\ref{fig:dipole} shows the induced dipole moment as a function of time for $N=30$ cycles and the corresponding electric field.
As seen, the magnitude of the dipole predicted by TDLPT is slightly smaller than that obtained from the reference solution.
At the peak intensity, where the absolute deviation is maximal, the TDLPT result differs by only $1\%$ from the full solution of the TDSE.
A similar agreement was observed for all optical-cycle numbers considered.

\subsubsection{Second-Order Correction and the Structure of Higher-Order Terms}
Similarly to \eqref{eq:factorization}, we assume the following natural factorization for the second correction
\begin{equation}
	\Phi_2(r,z,t)
	=\frac{\e^r}{r}\left(\varPhi_{2,0}(r,t)+\frac{z^2}{r^2}\varPhi_{2,2}(r,t)\right),
\end{equation}
where the functions $\varPhi_{2,0}$ and $\varPhi_{2,2}$ obey the following equations
\begin{equation}\label{eq:s}
	\ii\,\pa_{t}\,\varPhi_{2,0}
	=-\frac{1}{2}\,\pa_{r}^2\,\varPhi_{2,0}+\left(-\frac{1}{r}+\frac{1}{2}\right)\varPhi_{2,0}-\frac{\varPhi_{2,2}}{r^2}-\frac{e^r\varPhi_{1,1}^2}{2 r^3},
\end{equation}
and
\begin{equation}\label{eq:d}
	\ii\,\partial_{t}\,\varPhi_{2,2}
	=-\frac{1}{2}\,\partial_{r}^2\,\varPhi_{2,2}+\left(-\frac{1}{r}+\frac{3}{r^2}+\frac{1}{2}\right)\varPhi_{2,2}+e^r\left[\frac{\varPhi_{1,1}(\varPhi_{1,1}+\partial_r\varPhi_{1,1})}{r^2}-\frac{(\varPhi_{1,1}+\partial_r\varPhi_{1,1})^2}{2r}\right].
\end{equation}
Equation \eqref{eq:s} can be identified with the TDSE governing the evolution of an $s$~state, while \eqref{eq:d} describes the evolution of a $d$~state.
This identification is consistent with the dipole selection rule $\Delta\ell=\pm 1$ for hydrogen in linearly polarized light \cite{BetheSalpeter57} at an intermediate $p$~state generated by single-photon absorption from the ground state.
In contrast to standard TDPT, the present framework encodes transitions in the components $\varPhi_{2,0}$ and $\varPhi_{2,2}$ of the phase, rather than in the wave function itself.

\begin{figure}[h]
	\includegraphics[scale=1.5]{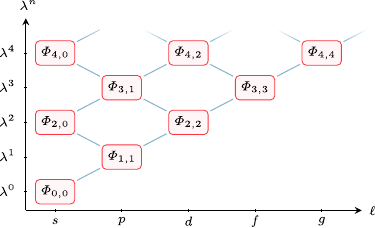}
	\caption{Diagrammatic representation of the terms contributing to the correction $\Phi_{2n+\sigma}$.
		We define $\varPhi_{0,0}:=\Phi_0$ and use spectroscopic notation to indicate the angular momentum.}\label{fig:selection_rules}
\end{figure}

From the first two corrections, we infer the following general structure:
\begin{equation}\label{eq:structure}
	\Phi_{2n+\sigma}(r,z,t)
	=\frac{z^\sigma\e^{r}}{r^{\sigma+1}}\sum_{k=0}^n\left(\frac{z}{r}\right)^{2k}\varPhi_{2n+\sigma,2k+\sigma}(r,t),\quad
	\sigma
	=0,1.
\end{equation}
Remarkably, the polynomial structure of the corrections in the variable $z$ coincides with the one obtained from time-independent LPT in the case of a static field~\cite{Turbiner1983Hydrogen}.
Diagrammatically, figure~\ref{fig:selection_rules} encodes the terms contributing to the correction \eqref{eq:structure}, while reflecting the well-known selection rules of the hydrogen atom.

In \eqref{eq:structure}, each function $\varPhi_{2n+\sigma,2k+\sigma}$ evolves according to the radial TDSE-like equation with orbital angular momentum $\ell=2k+\sigma$ 
\begin{align}
	\ii\,\partial_{t}\,\varPhi_{2n+\sigma,2k+\sigma}
	&=-\frac{1}{2}\,\partial_{r}^2\,\varPhi_{2n+\sigma,2k+\sigma}+\left(-\frac{1}{r}+\frac{(2k+\sigma)(2k+\sigma+1)}{2r^2}+\frac{1}{2}\right)\varPhi_{2n+\sigma,2k+\sigma}\nonumber\\
	&\qquad+\text{source}(r,t),\quad
	k=0,\ldots,n,
\end{align}
with a source term determined by previously calculated corrections, accompanied by
\begin{equation}
	\varPhi_{2n+\sigma,2k+\sigma}(r,0)
	=0,
	\quad\text{and}\quad
	\varPhi_{2n+\sigma,2k+\sigma}(0,t)
	=\varPhi_{2n+\sigma,2k+\sigma}(\infty,t)
	=0.
\end{equation}
As a final remark, we point out that the same form of the correction \eqref{eq:structure} arises for any $\hat{H}_0$ of the form \eqref{eq:choice}, with an arbitrary\footnote{In order to satisfy the same (asymptotic) boundary conditions \eqref{eq:boundary}, one needs to restrict $V(r)\sim 1/r^{\beta}$ with $\beta\le 2$ as $r\rar0$, a typical situation in physics.} radial potential $V(r)$ in place of the Coulomb potential $V(r)=-1/r$.
In particular, this includes model potentials, as commonly used in the celebrated single-active-electron approximation (see, e.g. \cite{RevModPhys.81.163}).
The latter approach has been widely used to simulate the quantum dynamics of multi-electron atomic systems, including alkali atoms \cite{Li} (e.g. sodium and rubidium), helium, and quasi-two-electron atoms such as beryllium and magnesium \cite{Reiff_2020}.

\section{Conclusions and Outlook}
\label{Sec:Conclusions}
In this work, we have developed the time-dependent extension of LPT in which the wave function is expressed in exponential form, with its argument expanded as a power series in the coupling constant.
This extension is carried out in a systematic and rigorous manner assuming that the system is initially in the unperturbed nondegenerate ground state.

Within the framework of TDLPT, the key features of the original time-independent formulation are retained, most notably the appearance of perturbative corrections in closed integral form, whose evolution is governed by a gauge-rotated free time-evolution operator. 
In this approach, dynamic energy shifts arise naturally in integral form and admit a transparent interpretation as time averages calculated via pseudopotentials.
In contrast to standard formulations of TDPT based on the Dyson series, TDLPT provides a direct route to computing these shifts for general time-dependent perturbations without summing over the whole family of unperturbed states; see \eqref{eq:standard}.

We have applied the method to two physically relevant systems: the harmonic oscillator and the hydrogen atom, both subjected to a time-dependent external field.
In the former case, TDLPT reproduces the exact wave function, thereby providing a proof of principle of the method.
In the latter, the system is analyzed using both analytical and numerical methods.
Moreover, we showed that the selection rules of the hydrogen atom can be inferred from the structure of the phase corrections.
Remarkably, the same structure of the correction terms extends to systems governed by arbitrary radial potentials.

The present work establishes a basis for further investigations in perturbative regimes within the single-active-electron approximation, including the strong-coupling regime via the Symanzik scaling.
It also opens the possibility of deriving analytical expressions for phase corrections and experimentally relevant observables --- such as AC Stark shifts, polarizabilities, and photoionization time delays --- in situations where standard perturbative approaches do not readily yield closed-form results.

We conclude by making two final remarks.
\begin{enumerate}
	\item The formulation of TDLPT developed here assumes that the system is initially in the unperturbed ground state.
		Extending the method to excited states remains an open problem, mainly due to the treatment of nodal structures.
		This difficulty is already present in LPT, where also no general solution is currently available.
		A possible route could be to factor out the known nodal structure of an unperturbed excited state and apply the logarithmic expansion only to the remaining non-vanishing factor.
	\item As in its time-independent counterpart, the method is expected to be most useful for low-dimensional systems or symmetry-reduced multidimensional problems.
		For genuinely multidimensional non-separable systems, one has to solve the resulting hierarchy of PDEs numerically, which is computationally demanding.
		However, we expect the time-dependent extension to be applicable whenever LPT itself provides a suitable framework for a given quantum system.
		The example of the hydrogen atom presented in section~\ref{sec:Hydrogen} supports this claim.
\end{enumerate}
A detailed treatment of both points is deferred to future work.

\section*{Acknowledgments}
The work of J.C.d.V. was partially supported by the UNAM-PAPIIT grant~IA100926.
P.B. was supported by the Polish NCN project no.~2022/45/B/ST1/00135.
The work of K.K. was funded by the Polish NCN project no. 2019/34/E/ST1/00390.
J.C.d.V. is grateful to the University of Gda\'nsk for its hospitality, where the main part of this work was carried out.
The authors are grateful to Prof. Klaus Bartschat for providing the computational resources that enabled the full solution of the TDSE. J.C.d.V. thanks Abraham Camacho for useful comments on the pulses used in this work.
We also thank the referees for their thorough comments, which helped us to improve the presentation of this work.

\bibliography{references}
\newpage

\appendix
\section{Gauge-Rotated Representation of the Semigroup}\label{sec:representation}
In this appendix, we show that the operator $\hat{\mathcal L}$ generates a strongly continuous unitary group on a suitable weighted Hilbert space.
Consider
\begin{equation}
	L^2(\Omega,\psi_0^2\,\mathrm{d}x)
	:=\left\{h\colon\Omega\to\mathbb{C}\,\mid\,\int_\Omega|h(x)|^2\psi_0(x)^2\,\mathrm{d}x<\infty\right\},
\end{equation}
equipped with the inner product
\begin{equation}
	\langle f\mid g\rangle_{\psi_0^2}
	:=\int_\Omega f(x)\overline{g(x)}\,\psi_0(x)^2\,\mathrm{d}x.
\end{equation}
Since the ground-state wave function $\psi_0$ is real-valued and nodeless, multiplication by $\psi_0$ defines the map
\begin{equation}
	M\colon L^2(\Omega,\psi_0^2\,\mathrm{d}x)\to L^2(\Omega),\quad
	(Mh)(x)
	=\psi_0(x)h(x).
\end{equation}
Moreover,
\begin{equation}
	\|Mh\|_{L^2(\Omega)}^2
	=\int_\Omega|h(x)|^2\psi_0(x)^2\,\mathrm{d}x
	=\|h\|_{L^2(\Omega,\psi_0^2\,\mathrm{d}x)}^2,
\end{equation}
which shows that $M$ is a unitary operator from $L^2(\Omega,\psi_0^2\,\mathrm{d}x)$ onto $L^2(\Omega)$.
In particular, the relation \eqref{eq:formula_LR} can be rewritten as
\begin{equation}
	\hat{\mathcal L}
	=M^{-1}\left(\hat H_0-E_0\right)M,
\end{equation}
which shows that $\hat{\mathcal L}$ is unitarily equivalent to the shifted Hamiltonian $\hat H_0-E_0$.

Since $\hat H_0$ is self-adjoint, Stone's theorem \cite{Pazy1983} implies that $\e^{-\ii t(\hat H_0-E_0)}$ forms a strongly continuous unitary group on $L^2(\Omega)$.
Consequently,
\begin{equation}
	\e^{-\ii t\hat{\mathcal L}}
	=\e^{\phi_0}\e^{-\ii t(\hat H_0-E_0)}\e^{-\phi_0}
	=M^{-1}\e^{-\ii t(\hat H_0-E_0)}M
\end{equation}
defines a strongly continuous unitary group on $L^2(\Omega,\psi_0^2\,\mathrm{d}x)$.
In particular,
\begin{equation}
	\|\e^{-\ii t\hat{\mathcal L}}h\|_{L^2(\Omega,\psi_0^2\,\mathrm{d}x)}
	=\|h\|_{L^2(\Omega,\psi_0^2\,\mathrm{d}x)}
\end{equation}
for all $h\in L^2(\Omega,\psi_0^2\,\mathrm{d}x)$ and all $t\in\mathbb{R}$.

We emphasize that this weighted space is the natural functional setting for the correction functions.
For instance, as can be seen from \eqref{eq:Phi_1}, the first correction term in the harmonic-oscillator case exhibits linear growth in $x$.
Hence $\Phi_1\notin L^2(\mathbb R)$, whereas
\begin{equation}
	\int_{\mathbb R}|\Phi_1(x,t)|^2\psi_0(x)^2\,\mathrm{d}x
	<\infty
\end{equation}
because $\psi_0$ is Gaussian.

\section{Connection to Standard TDPT}\label{app:connection}
Standard TDPT is formulated in terms of the Dyson series, which can be obtained via a bootstrapping construction based on Duhamel's formula applied to the TDSE.
Within the present approach, such an expansion is recovered by a power series of \eqref{eq:ExpRep} in the coupling parameter $\lambda$, thereby providing a direct connection between the conventional perturbative treatment and the TDLPT framework.

According to Duhamel's formula applied to \eqref{eq:main}, the time-evolved wave function takes the implicit form
\begin{equation}\label{eq:Dyson}
	\psi(x,t)
	=\psi_{0}(x)\,\e^{-\ii E_0 t}-\ii\,\lambda\int_{0}^{t}\e^{-\ii(t-s)\hat H_{0}}\,V_{\mathrm{int}}(x,s)\,\psi(x,s)\,\mathrm{d}s.
\end{equation}
We proceed by \emph{bootstrapping}; i.e., we insert the expression on the right-hand side of \eqref{eq:Dyson} into $\psi(x,s)$ under the integral, which yields
\begin{equation}
	\psi(x,t)
	=\psi_{0}(x)\,\e^{-\ii E_0 t}-\ii\,\lambda\int_{0}^{t}\e^{-\ii(t-s)\hat H_{0}}\,V_{\mathrm{int}}(x,s)\,\e^{-\ii sE_0}\psi_{0}(x)\,\mathrm{d}s+\mathcal{O}(\lambda^{2}).
\end{equation}
Higher-order terms in $\lambda$ are obtained by repeating this procedure and result in the following expression for the solution:
\begin{align}\label{eq:bootstrap}
	\psi(x,t)
	&=\psi_{0}(x)\,\e^{-\ii E_0 t}+\sum_{n=1}^\infty(-\ii\,\lambda)^n\int_0^t\int_0^{s_1}\cdots\int_0^{s_{n-1}}\Big[\e^{-\ii(t-s_1)\hat H_0}V_{\mathrm{int}}(x,s_1)\notag\\
	&\qquad\e^{-\ii(s_1-s_2)\hat H_0}V_{\mathrm{int}}(x,s_2)\cdots V_{\mathrm{int}}(x,s_n)\e^{-\ii s_nE_0}\psi_0(x)\Big]\mathrm{d}s_n\cdots\mathrm{d}s_1,
\end{align}
known as the Dyson series\footnote{The Dyson series arises as the Neumann-series expansion of the Duhamel integral equation, with time ordering $0\le s_n\le\cdots\le s_1\le t$ encoded in the nested integration limits.}.

In the next step, we now establish a counterpart formula via a direct expansion of the exponential in the ansatz \eqref{eq:ExpRep}.
Based on the expansion of the exponential series, we obtain
\begin{align}
	\psi(x,t)
	&=\e^{\sum_{j=0}^\infty\lambda^j\Phi_j(x,t)}
	=\e^{-\phi_0(x)-\ii E_0t}\prod_{j=1}^\infty\e^{\lambda^j\Phi_j(x,t)}\notag\\
	&=\e^{-\phi_0(x)-\ii E_0t}\prod_{j=1}^\infty\sum_{k_j=0}^\infty\frac{(\lambda^j\Phi_j(x,t))^{k_j}}{k_j!}.
\end{align}
Upon collecting powers of $\lambda$, the previous formula can be rewritten as
\begin{align}\label{eq:expansion_of_ansatz}
	\psi(x,t)
	&=\e^{-\phi_0(x)-\ii E_0 t}\sum_{k_1=0}^\infty\sum_{k_2=0}^\infty\cdots\lambda^{\sum_{j=1}^\infty jk_j}\prod_{j=1}^\infty\frac{\Phi_j(x,t)^{k_j}}{k_j!}\notag\\
	&=\e^{-\phi_0(x)-\ii E_0 t}\sum_{n=0}^\infty\lambda^n\sum_{k\in\Omega_n}\prod_{j=1}^n\frac{\Phi_j^{k_j}}{k_j!},
\end{align}
where we introduced the multi-index set $\Omega_n:=\{k\in\mathbb{N}^n\,\colon\,\sum_{j=1}^njk_j=n\}$.
A direct comparison of the bootstrap formula \eqref{eq:bootstrap} with \eqref{eq:expansion_of_ansatz} yields the following relation for the coefficient functions of $\lambda^n$, $n\ge 1$:
\begin{align}
	\e^{-\phi_0(x)-\ii E_0 t}\sum_{k\in\Omega_n}\prod_{j=1}^n\frac{\Phi_j^{k_j}}{k_j!}
	&=(-\ii)^n\int_0^t\int_0^{s_1}\cdots\int_0^{s_{n-1}}\Big[\e^{-\ii(t-s_1)\hat H_0}V_{\mathrm{int}}(x,s_1)\\
	&\qquad\e^{-\ii(s_1-s_2)\hat H_0}V_{\mathrm{int}}(x,s_2)\cdots V_{\mathrm{int}}(x,s_n)\e^{-\ii s_n\hat H_0}\psi_0(x)\Big]\mathrm{d}s_n\cdots\mathrm{d}s_1.\notag
\end{align}
While the case $n=1$ is verified from a simple calculation, $n>1$ establishes non-trivial relationships between the respective correction terms.
We note that the above expansions need to be understood in a formal way.
In fact, the rigorous analysis of their convergence is an object of major concern and a challenging mathematical problem which is left for future research.
We conclude this subsection by noting that due to the presence of a time-dependent potential in the TDSE \eqref{eq:main} its solution $\psi$ can also be expressed in terms of the celebrated \emph{Magnus} or \emph{Fer expansions} \cite{MR1883629,MR1866600,MR2494199,MR4868053}.
The latter can be reordered in powers of $\lambda$ to obtain alternative non-trivial relationships between the correction terms.
However, the rigorous convergence analysis of Magnus and Fer expansions remains an active topic of research in the scientific community.

\section{Dynamic Energy Shift for the Harmonic Oscillator}\label{app:Stark}
In this appendix, we provide details for the calculation of the time-averaged energy shift.
Performing the time integration in equation~\eqref{eq:adiabatic} yields
\begin{equation}
	\Phi_1(x,t)
	=-\ii\,x\,W(t),
	\quad\text{where}\quad
	W(t)
	=\frac{\ii\cos(\omega t)+\omega\sin(\omega t)}{\omega^{2}-1}.
\end{equation}
Here, the limit $\epsilon\to 0^{+}$ has already been taken.
It then follows that the instantaneous energy shift \eqref{eq:instantaneous_energy} is given by
\begin{equation}
	\langle Q_2(x,t)\rangle_{\psi_0}
	=-\frac{1}{2}\bigl[\partial_x\Phi_1(x,t)\bigr]^2
	=\frac{1}{2}\,W(t)^2.
\end{equation}
We now evaluate the time average over one period of the driving field ($T=2\pi/\omega$, $t_0=T/2$; see \eqref{eq:average_energy_correction}) to obtain a real-valued energy shift:
\begin{equation}
	\overline{E}_2
	=\overline{\langle Q_2(x,t)\rangle}_{\psi_0}
	=\frac{1}{2\,T}\int_0^{T}\!\!\!W(t)^2\,\mathrm{d}t
	=\frac{1}{4(\omega^{2}-1)}.
\end{equation}
In particular, $\overline{E}_2$ is independent of $t_0$ and $T$ due to the periodicity of $W(t)$.

\section{Hybrid Coordinates}\label{app:coordinates}
\begin{figure}[h!]
	\includegraphics[scale=1]{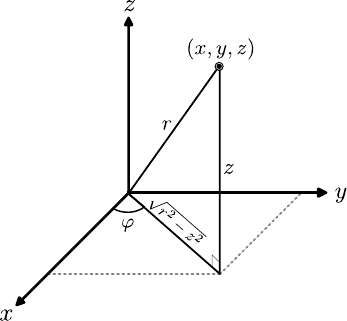}
	\caption{Graphical representation of the hybrid coordinates $\{r,\varphi,z\}$ in Cartesian space.
	\label{fig:hybrid}}
\end{figure}

Hybrid coordinates may be regarded as a combination of three-dimensional spherical and cylindrical coordinate systems.
They are related to Cartesian coordinates according to\footnote{$r\in[0,\infty)$, $\varphi\in[0,2\pi)$, and $z\in[-r,r]$.}
\begin{equation}
	x
	=\sqrt{r^2-z^2}\,\cos\varphi,\qquad
	y
	=\sqrt{r^2-z^2}\,\sin\varphi,
\end{equation}
while the $z$-coordinate remains unchanged.
A visualization of the coordinate geometry is shown in figure~\ref{fig:hybrid}.


A straightforward calculation shows that the Laplacian is given by
\begin{equation}\label{eq:Laplacian}
	\Delta
	=\partial_x^2+\partial_y^2+\partial_z^2
	=\pa_r^2+\frac{2}{r}\pa_r+\frac{2z}{r}\pa_{rz}+\pa_z^2+\frac{1}{r^2-z^2}\pa_{\varphi}^2,
\end{equation}
while the product of the gradients results in 
\begin{equation}\label{eq:grad_dot_grad}
	\nabla f\cdot\nabla g
	=\pa_rf\cdot\pa_rg+\frac{z}{r}(\pa_rf\cdot\pa_zg+\pa_zf\cdot\pa_rg)+\pa_zf\cdot\pa_zg+\frac{\pa_\varphi f\cdot\pa_\varphi g}{r^2-z^2},
\end{equation}
where $f$ and $g$ are two arbitrary functions.
Finally, we remark that the volume element $\mathrm{d}V$ can be expressed as
\begin{equation}
	\mathrm{d}V
	=r\,\mathrm{d}r\,\mathrm{d}\varphi\,\mathrm{d}z.
\end{equation}	
\end{document}